\documentclass[11pt]{article}
\usepackage{amssymb}
\usepackage{amsmath}
\usepackage[affil-it]{authblk}

\usepackage{fancyhdr}
\fancypagestyle{plain}{%
        \fancyhead[R]{ITEP-TH-28/13}
        
}

\begin{document}
\author{{Oleg Dubinkin\thanks{dubinkin@itep.ru}}\\{\normalsize\rm Institute for Theoretical and Experimental Physics, 117259 Moscow, Russia}\\ \rm \normalsize and\\  {\normalsize\rm Moscow Institute of Physics and Technology, 141700 Dolgoprudny, Russia}}
\date{}
\title{\Huge On the Virasoro constraints for torus knots}

\maketitle
\begin{abstract}
We construct Virasoro algebra of differential operators for the matrix model for torus knots. These operators generate various relations between Wilson loops. Then we discuss the constructed operators and corresponding relations in the stability limit. A series of examples is given. 
\end{abstract}
\section{Introduction}

\qquad Matrix models provide a rich set of approaches to physical systems and quantities (see \cite{faces} for review). Constructing a fundamental Virasoro algebra for considered matrix model is one of the most useful methods. This algebra is very fruitful by itself, as it allows us to generate various relations between quantities represented in this model. One can go further, for example, constructing a spectral curve using Virasoro algebra. But the first step is to derive one, which is the subject of this paper. We will study the matrix model, that describes Wilson loops along torus knots. To obtain Virasoro algebra we will use a standard procedure. First, one should input a generating function for desirable quantities into matrix integral and obtain a Ward identity for this new integral. The second step is to construct differential operators, that can produce such Ward identities being applied to the matrix integral with generating function. This procedure is not always possible, so it can be necessary to input several generating functions in the integrand. We apply this technique to the considered matrix integral and this paper results in generators of Virasoro algebra for torus knots (\ref{win}) and identities between Wilson loops, produced by generators of this algebra.

\quad This paper is organized as follows. Section 2 reviews the general algorithm for the matrix integral of Chern-Simons partition function and obtains trivial differential equation for it as an example. In section 3, we will discuss different approaches to the matrix integral for torus knots and obtain two different representations of Virasoro algebra that corresponds to different generating functions. Section 4 discusses the so-called stability limit, where we will see decomposition of Virasoro generators into three independent sets of differential operators. Finally, section 5 describes a series of examples.
\section{Matrix integral and Ward identities for torus knot}

$\qquad$Let us consider a matrix integral expression for Wilson loops in Chern-Simons theory with gauge group G=SU(N), written for the case of torus knots. This expression was first obtained for the case of SU(2) in \cite{SU2} and then extended to simply-laced groups in \cite{SimLac} (for additional information see \cite{JR, TierzCS, Mar}). Also recently, this matrix integral representation was derived by localization procedure \cite{Beasley, Kallen}. The result obtained in these papers reads:
\begin{equation}
W_R(\mathcal{K}_{P,Q})=\frac{1}{Z(\mathcal{K}_{P,Q})}\int du e^{-\frac{u^2}{2g_s}}\prod_{\alpha>0}4\sinh(\frac{u\cdot \alpha}{2P})\sinh(\frac{u\cdot \alpha}{2Q})\chi_R(e^u)
\end{equation}
In this expression,
\begin{equation}
\label{partition}
Z(\mathcal{K}_{P,Q})=\int du e^{-\frac{u^2}{2g_s}}\prod_{\alpha>0}4\sinh(\frac{u\cdot \alpha}{2P})\sinh(\frac{u\cdot \alpha}{2Q})
\end{equation}
and $\chi_R(e^u)$ are the characters of G, $g_s=\frac{PQ\cdot 2\pi i}{\kappa+h}$, $\kappa$ is the level of the theory, $h$ is the dual Coxeter number of G, $u$ is an element in $\Lambda_w \otimes \mathbb{R}$ and $\alpha$ is a set of positive simple roots of G. A torus knot is specified by a pair of coprime integers $P$ and $Q$. In this section we will work with the integral (\ref{partition}) to develop the necessary technique and show the simplest example of deriving Ward identity and the corresponding differential operator. Using a well-known fact that the positive roots can be written in an orthonormal basis $\{e_k\}_{k=1,...,N}$ as
$$
\alpha_{kl}=e_k-e_l, 1\leqslant k<l\leqslant N
$$
and that $u=\sum_k \lambda_k e_k$, one can rewrite the integral in a form, which is more appropriate for further calculations:
\begin{equation}
Z(\mathcal{K}_{Q,P})=\int \prod_{i=1}^Nd\lambda_i e^{-\frac{\lambda_i^2}{2g_s}}\prod_{k<l}4\sinh(\frac{\lambda_k-\lambda_l}{2P})\sinh(\frac{\lambda_k-\lambda_l}{2Q})
\end{equation}
The following considerations are very similar to those in \cite{Lagapp}. To obtain Ward identities, one should perform the next substitution:
\begin{equation}
\tilde{\lambda}_i=\lambda_i+\epsilon f_i, \quad \lambda_i=\tilde{\lambda}_i-\epsilon f_i
\end{equation}
Considering $\epsilon$ an infinitesimal we can decompose integrals over each $\lambda_i$ in $Z_{P,Q}$ as follows: $\int f(\lambda)d\lambda=\int f(\tilde{\lambda})d\tilde{\lambda}+\epsilon\int g(\tilde{\lambda})d\tilde{\lambda}+o(\epsilon^2)$. In the first order of $\epsilon$ this gives us: $\int g(\lambda)d\lambda=0$. Performing such procedure for each $\lambda_i$ in $Z_{P,Q}$ one obtains the following Ward identity:
\begin{equation}
\begin{split}
\begin{gathered}
\int \prod_{i=1}^Nd\lambda_i \left(\sum_{i=1}^{N}(\frac{\partial f_i}{\partial \lambda_i}-\frac{\lambda_i f_i}{g_s}) + \frac{f_k-f_l}{2Q}\coth(\frac{\lambda_k-\lambda_l}{2Q}) + \frac{f_k-f_l}{2P}\coth(\frac{\lambda_k-\lambda_l}{2P})\right)\times\\\times e^{-\frac{\lambda_i^2}{2g_s}}\prod_{k<l}4\sinh(\frac{\lambda_k-\lambda_l}{2P})\sinh(\frac{\lambda_k-\lambda_l}{2Q})=0
\end{gathered}
\end{split}
\end{equation}
Now we put $f_i=\lambda_i$ and, to shorten the expression, we replace integrals with brackets (this notation will be used throughout this paper):
\begin{equation}
\Big\langle \sum_{i=1}^{N}(1-\frac{\lambda_i^2}{g_s}) \Big\rangle+\Big\langle \frac{\lambda_k-\lambda_l}{2Q}\coth(\frac{\lambda_k-\lambda_l}{2Q}) \Big\rangle+\Big\langle \frac{\lambda_k-\lambda_l}{2P}\coth(\frac{\lambda_k-\lambda_l}{2P}) \Big\rangle=0
\end{equation}
The resulting identity can be obtained by applying the following differential operator to $Z_{P,Q}$: 
\begin{equation}
\widehat{L}=N-P\frac{\partial}{\partial P}-Q\frac{\partial}{\partial Q}, \qquad \widehat{L}Z_{P,Q}=0.
\end{equation}
The last equation simply states that $Z_{P,Q}$ is a homogeneous function of degree $N$ with respect to variables $P$ and $Q$. This statement is easily verified by taking the integral.

\section{Elements of Virasoro algebra}

$\qquad$Now we will take a look at the observables in Chern-Simons theory - Wilson loops. In terms of this matrix integral, Wilson loops for torus knots $\mathcal{K}_{Q,P}$ are the averages of the characters of G, which can be expressed through Schur polynomials by the Weyl determinant formula as $\chi_R(t)=\det(s_{\lambda_i-i+j}(t))$ (for additional information see \cite{Morunit}):
\begin{equation}
W_R(\mathcal{K}_{P,Q})=\frac{1}{Z_{P,Q}}\int \prod_{i=1}^Nd\lambda_i e^{-\frac{\lambda_i^2}{2g_s}}\prod_{k<l}4\sinh(\frac{\lambda_k-\lambda_l}{2P})\sinh(\frac{\lambda_k-\lambda_l}{2Q})\chi_R(e^\lambda)=\Big\langle\chi_R(e^\lambda)\Big\rangle
\end{equation}
For further calculations it is convenient to perform the following change of variables \cite{Mar,TierzSoft,TierzSchur}:
\begin{equation}
\lambda_i=PQ\log x_i
\end{equation}
The expression for a Wilson loop will then take the following form:
\begin{equation}
\begin{gathered}
\label{wilson_log}
W_R(\mathcal{K}_{P,Q})=\frac{(PQ)^N}{Z_{P,Q}}\int_{\mathbb{R}_+^N} \prod_{i=1}^Ndx_i \exp\left(-\frac{PQ(\log x_i)^2}{2\hat{g}_s}-\left(\frac{P+Q}{2}(N-1)+1\right)\log x_i\right)\times\\ \times\prod_{k<l}(x_k^P-x_l^P)(x_k^Q-x_l^Q)\chi_R(x)
\end{gathered}
\end{equation}
where $\hat{g}_s=\frac{2\pi i}{\kappa+h}$. Let us replace the characters by their generating function: 
\begin{equation}
\sum_R\chi_R(t)\chi_R(x)=\exp\left(\sum_{k=1}^\infty t_kp_k\right), \quad \textup{where}\quad p_k=\sum_{i=1}^Nx_i^k
\end{equation}
and the resulting integral will be: 
\begin{equation}
W=\sum_R\chi_R(t)W_R(\mathcal{K}_{P,Q})
\end{equation} 
The next step consists of obtaining Ward identities for $W$, performing a change of variables as in the previous section. Then we will construct differential polynomials with respect to $\{\frac{\partial}{\partial t_k}\}$ that generate these identities. Applying them to $W$ and fixing the infinite set of parameters $\{t_k\}$, one will obtain various relations between Wilson loops for the same knot in different representations. Unfortunately, since logarithm can't be represented as a power series of $x_i$, that converges on whole positive semi-axis, we can't construct a term with $\log(x_i)$ in upcoming Ward identity (\ref{wardlog}), using only differentials with respect to $t_k$'s. The most evident way to solve this problem comprises inserting another generating function into the integral:
\begin{equation}
\exp\left(\sum_{k=1}^\infty l_k\left(\sum_{i=1}^Nx_i^k\log x_i \right)\right)
\end{equation}
This functions also should be interpreted as character decomposition. From now on we will be considering the following integral:
\begin{equation}
\label{tildeW}
\begin{split}
\widetilde{W}=\frac{(PQ)^N}{Z_{P,Q}}\int_{\mathbb{R}_+^N} \prod_{i=1}^Ndx_i \exp\left(-\frac{PQ(\log x_i)^2}{2\hat{g}_s}-\left(\frac{P+Q}{2}(N-1)+1\right)\log x_i\right)\times\\ \times\prod_{k<l}(x_k^P-x_l^P)(x_k^Q-x_l^Q)\exp\left(\sum_{k=1}^\infty t_kp_k\right)\exp\left(\sum_{k=1}^\infty l_k\left(\sum_{i=1}^Nx_i^k\log x_i \right)\right)
\end{split}
\end{equation}
Repeating the procedure described in the previous section, we obtain Ward identity:
\begin{equation}	
\label{wardlog}
\begin{gathered}
\Big\langle \sum_{i=1}^{N}\left(\frac{\partial f_i}{\partial x_i}-\frac{PQ}{\hat{g}_s}\frac{f_i\log x_i}{x_i}-\left(\frac{P+Q}{2}(N-1)+1\right)\frac{f_i}{x_i}\right) \Big\rangle+\\+\Big\langle Q\sum_{k<l}\frac{x_k^{Q-1}f_k-x_l^{Q-1}f_l}{x_k^Q-x_l^Q} \Big\rangle+\Big\langle P\sum_{k<l}\frac{x_k^{P-1}f_k-x_l^{P-1}f_l}{x_k^P-x_l^P} \Big\rangle+\\+\Big\langle \sum_{k=1}^{\infty}kt_k \left(\sum_{i=1}^Nx_i^{k-1}f_i\right) \Big\rangle+\Big\langle \sum_{k=1}^{\infty}l_k \left(\sum_{i=1}^Nf_ix_i^{k-1}(1+k\log x_i)\right) \Big\rangle=0
\end{gathered}
\end{equation}
Note that we have omitted generating functions inside the brackets to make the expression shorter. This notation with skipped generating functions will be used throughout this paper. Now we have to choose actual form of the deformations $f_i$'s. The most common way to do so is to put $f_i=x_i^{nPQ+1}$. The resulting identities can be obtained by applying the following differential operators to $\widetilde{W}$:
\begin{equation}
\label{opM}
\begin{split}
\widehat{M}_{nPQ}=\left( nPQ-(P+Q)(N-1) \right)\frac{\partial}{\partial t_{nPQ}}-\frac{PQ}{\hat{g}_s}\frac{\partial}{\partial l_{nPQ}}+P\sum_{a+b=nQ}\frac{\partial}{\partial t_{aP}}\frac{\partial}{\partial t_{bP}}+\\+Q\sum_{a+b=nP}\frac{\partial}{\partial t_{aQ}}\frac{\partial}{\partial t_{bQ}}+\sum_{k=1}^{\infty}kt_k\frac{\partial}{\partial t_{nPQ+k}}+\sum_{k=1}^{\infty}l_k\left( \frac{\partial}{\partial t_{nPQ+k}}+k\frac{\partial}{\partial l_{nPQ+k}} \right)
\end{split}
\end{equation}
This operators form the Virasoro algebra. Applying them to $\widetilde{W}$ and defining two infinite sets of parameters $\{t_k\}$ and $\{l_k\}$, one can generate identities between Wilson loops in different representations and averages of the form $\langle \chi_{R_1}(p_k) \chi_{R_2}(\tilde{p}_k) \rangle$, where $p_k=\sum_{i=1}^Nx_i^k$ and $\tilde{p}_k=\sum_{i=1}^Nx_i^k\log(x_i)$. For example, if we put all $t_k$'s and $l_k$'s equal to zero, we will obtain relations between average $\langle \sum x_i^{nPQ}\log (x_i) \rangle$ and Wilson loops, taken in different representations with the same number of boxes in corresponding Young tableau. For more details, see section 5.

\quad There is a way to rewrite this matrix model without the logarithmic term. This will allow us to obtain Virasoro algebra using only one generating function, which is our final goal in this section. Let us consider the following simple identity of integrals (as in \cite{Shamapp}, appx. B):
\begin{equation}
\label{complex}
\frac{1}{\sqrt{2\pi g}}\int_0^\infty dx\; x^d \exp \left(\frac{(\log x)^2}{2g}\right)=q^{(d+1)^2/2}=\frac{1}{2\pi}\oint_{|z|=1}dz\;z^d\;\theta_q(z)
\end{equation}
where $q=e^{g}$ and $\theta_q(z)=\sum_{k=-\infty}^\infty q^{k^2/2}z^k$ is Jacobi theta function. This allows us to rewrite the matrix integral for Wilson loops (\ref{wilson_log}) in the following way:
\begin{equation}
W_R(\mathcal{K}_{P,Q})=\frac{(2\pi \hat{g}_s PQ)^{N/2}}{Z_{P,Q}}\oint_{|z|=1} \prod_{i=1}^Ndz_i\;  z_i^{-\left(\frac{P+Q}{2}(N-1)+1\right)}\theta_q(z_i)\prod_{k<l}(z_k^P-z_l^P)(z_k^Q-z_l^Q)\chi_R(z)
\end{equation}
with $q=\exp(\frac{\hat{g}_s}{PQ})$. We will make a small shift of coupling constant $\hat{g}_s\to \hat{g}_s-\epsilon$, where $0<\epsilon<<1$. Since $\operatorname{Re}\hat{g}_s=0$ and $P$, $Q$ $\in \mathbb{Z}$ we get $|q|<1$. Also note that $|z_i|=1$. As usual, we will work with the modified integral, where characters are replaced with their generating function: 
\begin{equation}
\label{W}
W=\frac{(2\pi \hat{g}_s PQ)^{N/2}}{Z_{P,Q}}\oint_{|z|=1} \prod_{i=1}^Ndz_i\;  z_i^{-\left(\frac{P+Q}{2}(N-1)+1\right)}\theta_q(z_i)\prod_{k<l}(z_k^P-z_l^P)(z_k^Q-z_l^Q)\exp\left(\sum_{k=-\infty}^\infty t_kp_k\right)
\end{equation}
where $p_k=\sum_{i=1}^Nz_i^k$. Note that summation in generating function is taken over all integers including negative values. This means that we are actually using two generating functions for positive and negative values of $k$. Performing transformation of variables, as in the previous case:
\begin{equation}
\tilde{z}_i=z_i+\epsilon z_i^{nPQ+1}, \quad z_i=\tilde{z}_i-\epsilon \tilde{z}_i^{nPQ+1}
\end{equation}
we obtain the following identity for $W$ integral in the first order of $\epsilon$:
\begin{equation}
\begin{gathered}
\label{Ward}
\Big\langle\left(nPQ-\frac{P+Q}{2}(N-1)\right)\sum_{i=1}^Nz_i^{nPQ}\Big\rangle+\Big\langle \sum_{i=1}^Nz_i^{nPQ}\frac{\sum_{k=-\infty}^{\infty}kq^{k^2/2}z^k_i}{\theta_q(z_i)} \Big\rangle+\\+\Big\langle Q\sum_{k<l}\frac{z_k^{Q(nP+1)}-z_l^{Q(nP+1)}}{z_k^Q-z_l^Q} \Big\rangle+\Big\langle P\sum_{k<l}\frac{z_k^{P(nQ+1)}-z_l^{P(nQ+1)}}{z_k^P-z_l^P} \Big\rangle+\Big\langle \sum_{k=-\infty}^{\infty}kt_k \left(\sum_{i=1}^Nz_i^{nPQ+k}\right) \Big\rangle=0
\end{gathered}
\end{equation}
The second term in the above expression can be transformed as follows:
\begin{equation}
\Big\langle \sum_{i=1}^Nz_i^{nPQ}\frac{\sum_{k=-\infty}^{\infty}kq^{k^2/2}z^k_i}{\theta_q(z_i)} \Big\rangle=\Big\langle \sum_{i=1}^Nz_i^{nPQ}\frac{z_i\,\theta_q(z_i)'_{z_i}}{\theta_q(z_i)} \Big\rangle=\Big\langle \sum_{i=1}^Nz_i^{nPQ+1}(\log\theta_q(z_i))'_{z_i} \Big\rangle
\end{equation} 
We will use the next representation for Jacobi theta function, that will help us to rewrite the logarithmic term in a more convenient way:
\begin{equation}
\theta_q(z_i)=\prod_{k=1}^{\infty}(1-q^k)(1+q^k\frac{z_i}{\sqrt{q}})(1+q^k\frac{1}{z_i\sqrt{q}})
\end{equation}
\begin{equation}
\begin{gathered}
z_i(\log\theta_q(z_i))'_{z_i}=z_i\left(\sum_{k=1}^{\infty}\left(\log\left(1-q^k\right)+\log\left(1+q^k\frac{z_i}{\sqrt{q}}\right)+\log\left(1+q^k\frac{1}{z_i\sqrt{q}}\right)\right)\right)'_{z_i}=\\=z_i\sum_{k=1}^{\infty}\left(\frac{q^{k-\frac{1}{2}}}{1+z_iq^{k-\frac{1}{2}}}-\frac{1}{z_i}\frac{1}{1+z_iq^{-(k-\frac{1}{2})}}\right)
\end{gathered}
\end{equation}
Now, using that $|q|<1$ and $|z_i|=1$, we can decompose these two fractions into the Laurent power series:
\begin{equation}
\begin{gathered}
\label{log_dif}
\sum_{k=1}^{\infty}\left(\sum_{l=1}^{\infty}(-1)^{l+1}(z_iq^{k-\frac{1}{2}})^l-\sum_{l=1}^{\infty}(-1)^{l+1}(\frac{1}{z_i}q^{k-\frac{1}{2}})^l\right)=\\=\sum_{l=-\infty}^{-1}(-1)^l\left(\frac{1-q}{\sqrt{q}}z_i\right)^l-\sum_{l=1}^{\infty}(-1)^l\left(\frac{\sqrt{q}}{1-q}z_i\right)^l=\sum_{l=1}^{\infty}(-1)^{l}\left(\frac{\sqrt{q}}{1-q}\right)^{l}(z_i^{-l}-z_i^{l})
\end{gathered}
\end{equation}
Finally, we can write the second term in the Ward identity in a more appropriate form:
\begin{equation}
\begin{gathered}
\Big\langle \sum_{i=1}^Nz_i^{nPQ}\frac{\sum_{k=-\infty}^{\infty}kq^{k^2/2}z^k_i}{\theta_q(z_i)} \Big\rangle=\Big\langle \sum_{l=1}^{\infty}(-1)^{l}\left(\frac{\sqrt{q}}{1-q}\right)^{l}\left(z_i^{nPQ-l}-z_i^{nPQ+l}\right)\Big\rangle
\end{gathered}
\end{equation}
Thus we can finally write down differential operators corresponding to Ward identity (\ref{Ward}):
\begin{equation}
\begin{gathered}
\label{win}
\widehat{L}_{nPQ}=\left(nPQ-(P+Q)(N-1)\right)\frac{\partial}{\partial t_{nPQ}}+\sum_{l=1}^{\infty}(-1)^{l}\left(\frac{\sqrt{q}}{1-q}\right)^{l}\left(\frac{\partial}{\partial t_{nPQ-l}}-\frac{\partial}{\partial t_{nPQ+l}}\right)+\\+P\sum_{a+b=nQ}\frac{\partial}{\partial t_{aP}}\frac{\partial}{\partial t_{bP}}+Q\sum_{a+b=nP}\frac{\partial}{\partial t_{aQ}}\frac{\partial}{\partial t_{bQ}}+\sum_{k=-\infty}^{\infty}kt_k\frac{\partial}{\partial t_{nPQ+k}}
\end{gathered}
\end{equation}
As mentioned already, we have used two generating functions. First one is $\exp\left(\sum_{k=1}^\infty t_kp_k\right)$ which is the generating function for the characters of G. The second generating function is $\exp\left(\sum_{k=-\infty}^{-1} t_kp_k\right)$. It can be represented as a character decomposition, but in this case, variables $p_k=\sum_{i=1}^Nz_i^k$ in character functions have negative values of $k$'s. We denote characters with negative $k$'s as $\chi_R(p_-)$ and the character functions with positive $k$'s as $\chi_R(p_+)$. Considering the integral (\ref{W}), it is easy to see that $\langle\chi_R(p_-)\rangle_{P,Q}=\langle\chi_R(p_+)\rangle_{-P,-Q}=\langle\chi_R(p_+)\rangle_{P,Q}$. Thus the integral (\ref{W}) can be written as:
\begin{equation}
W=\sum_R\sum_{R'}\chi_R(t_+)\chi_{R'}(t_-)\langle\chi_R(p_+)\chi_{R'}(p_-)\rangle
\end{equation}
After we apply the differential operator (\ref{win}) to the expression above and put all $t_k$'s equal to zero, we obtain the relations of the form:
\begin{equation}
\sum_RA_R\langle\chi_R(p_-)\rangle+\sum_RB_R\langle\chi_R(p_+)\rangle=\sum_R(A_R+B_R)\langle\chi_R(p_+)\rangle=0
\end{equation}
where $A_R$ and $B_R$ are some constants.

\section{Stability limit}

$\quad$Now let us consider the stability limit, i.e.:
\begin{equation}
P\to \infty, \quad Q\to \infty,\quad \textup{while}\quad \frac{P}{Q}=\textup{Const}.
\end{equation}
In this and the following sections we will carry out calculations for $\widetilde{W}$ integral (\ref{tildeW}) and corresponding $\widehat{M}_{nPQ}$ operators (\ref{opM}), but all of the following conclusions can be easily rewritten for $W$ integral and $\widehat{L}$ operators (\ref{win}). First, our Ward identity (\ref{wardlog}) with $f_i=x_i^{nPQ+1}$ must be resolved in all orders of $P$ and $Q$. Thus we end up with three independent equations:
$$
\Big\langle \sum_{i=1}^{N}nPQ x_i^{nPQ}\Big\rangle-\Big\langle \sum_{i=1}^{N}\frac{PQ}{\hat{g}_s}x_i^{nPQ}\log x_i\Big\rangle=0
$$
\begin{equation}
\Big\langle \sum_{i=1}^{N}\frac{P+Q}{2}(N-1)x_i^{nPQ}) \Big\rangle=\Big\langle Q\sum_{k<l}\frac{x_k^{Q+nPQ}-x_l^{Q+nPQ}}{x_k^Q-x_l^Q} \Big\rangle+\Big\langle P\sum_{k<l}\frac{x_k^{P+nPQ}-x_l^{P+nPQ}}{x_k^P-x_l^P} \Big\rangle
\end{equation}
$$
\Big\langle \sum_{k=0}^{\infty}kt_k \left(\sum_{i=1}^Nx_i^{k+nPQ}\right) \Big\rangle+\Big\langle \sum_{k=0}^{\infty}l_k \left(\sum_{i=1}^Nx_i^{k+nPQ}(1+k\log x_i)\right) \Big\rangle=0
$$
And, correspondingly, every single Virasoro algebra generator factorizes into three simple operators: 
$$
\hat{e}_{n}=nPQ\frac{\partial}{\partial t_{nPQ}}-\frac{PQ}{\hat{g}_s}\frac{\partial}{\partial l_{nPQ}}
$$
\begin{equation}
\label{stabvir}
\hat{f}_{n}=-(P+Q)(N-1)\frac{\partial}{\partial t_{nPQ}}+P\sum_{a+b=nQ}\frac{\partial}{\partial t_{aQ}}\frac{\partial}{\partial t_{bQ}}+Q\sum_{a+b=nP}\frac{\partial}{\partial t_{aP}}\frac{\partial}{\partial t_{bP}}
\end{equation}
$$
\hat{h}_{n}=\sum_{k=0}^{\infty}kt_k\frac{\partial}{\partial t_{nPQ+k}}+\sum_{k=0}^{\infty}l_k\left( \frac{\partial}{\partial t_{nPQ+k}}+k\frac{\partial}{\partial l_{nPQ+k}} \right)
$$
Thus, in this limit, we got much simpler expressions for Ward identities and corresponding differential operators. As we will see, this simplifies things greatly when we want to get actual relations for Wilson loops.

\section{Trefoil example}

$\quad$Now let us consider a trefoil knot and generate relations between its Wilson loops, or HOMFLY polynomials, written in different representations. To do so, one should consider generalized matrix integral (\ref{tildeW}) and put $P=2$, $Q=3$. This matrix integral can be written as:
\begin{equation}
\label{ftildeW}
\widetilde{W}=\sum_{R}\sum_{R'}\chi_R(t_k)\chi_{R'}(l_k)\langle \chi_R(p_k) \chi_{R'}(\tilde{p}_k) \rangle
\end{equation}
where $p_k=\sum_{i=1}^Nx_i^k$ and $\tilde{p}_k=\sum_{i=1}^Nx_i^k\log(x_i)$. We are interested in $\langle \chi_R(p_k)\rangle$ as they appear to be $W_R(\mathcal{K}_{P,Q})$, this is why we should put all $l_k$'s and $t_k$'s equal to zero -- that will leave us only separate averages of $\chi_R(p_k)$ and $\chi_R(\tilde{p}_k)$ in (\ref{ftildeW}) and put generating functions to $1$ in Ward identities. Now let us write an exact form of $n=1$ element of Virasoro algebra for trefoil:
\begin{equation}
\begin{split}
\widehat{M}_{nPQ}=\widehat{M}_{6}=\left( 6-5(N-1) \right)\frac{\partial}{\partial t_{6}}-\frac{6}{\hat{g}_s}\frac{\partial}{\partial l_{6}}+2\sum_{a+b=3}\frac{\partial}{\partial t_{2a}}\frac{\partial}{\partial t_{2b}}+\\+3\sum_{a+b=2}\frac{\partial}{\partial t_{3a}}\frac{\partial}{\partial t_{3b}}+\sum_{k=0}^{\infty}kt_k\frac{\partial}{\partial t_{6+k}}+\sum_{k=0}^{\infty}l_k\left( \frac{\partial}{\partial t_{6+k}}+k\frac{\partial}{\partial l_{6+k}} \right)
\end{split}
\end{equation}
Next, one should apply this differential polynomial to $\widetilde{W}$ and put all $\{t_k\}$ and $\{l_k\}$ equal to zero, as was discussed above and take, for example, simplest case of $2\times2$ matrix:
\begin{equation}
\widehat{M}_{6}\widetilde{W}|_{t_k=l_k=0, \forall k}=0
\end{equation}
$$
\Updownarrow
$$
\begin{equation}
\begin{split}
\frac{6}{\hat{g}_s}\langle \sum_{i=1}^Nx_i^6\log x_i \rangle=5W_{(6)}-5W_{(5,1)}+2W_{(4,1,1)}-3W_{(2,1,1,1,1)}+\\+3W_{(1,1,1,1,1,1)}+3W_{(4,2)}-W_{(3,3)}-W_{(2,2,2)}+3W_{(2,2,1,1)}-2W_{(3,2,1)}
\end{split}
\end{equation}
In a considered case, $N=2$ and the average of $\sum x_i^6\log(x_i)$ will be:
\begin{equation}
\label{lognPQ}
\langle \sum_{i=1}^Nx_i^6\log x_i \rangle=48\pi \hat{g}_s^2\exp\left(\frac{37\hat{g}_s}{24}\right)\left(7-11e^{\hat{g}_s}-13e^{2\hat{g}_s}+17e^{5\hat{g}_s}\right)
\end{equation}
It is easy to see that in LHS of equation (\ref{lognPQ}) we always have a term proportional to $\langle \sum x_i^{nPQ}\log(x_i) \rangle$. That means one can easily rewrite this identity as the identity between Wilson loops in different representations for two different knots if it is possible to choose such $n$'s that the products of $n$, $P$ and $Q$ for this knots are equal. For example, we can compare Wilson loops for trefoil knot with $n=2$ and for torus knot with  $P=4$, $Q=3$, $n=1$, but the result is too overwhelming to present it here. 

\quad In the stability limit, as was discussed above, we get two following identities (the third one vanishes as we put $t_k$'s and $l_k$'s equal to zero) for Wilson loops, when $\frac{P}{Q}=\frac{3}{2}$:
\begin{equation}
\frac{6}{\hat{g}_s}\langle \sum_{i=1}^Nx_i^6\log x_i \rangle=W_{6}-W_{5,1}+W_{4,1,1}-W_{3,1,1,1}+W_{2,1,1,1,1}-W_{1,1,1,1,1,1}\equiv6\Big\langle \sum_{i=1}^{N} x_i^{6}\Big\rangle
\end{equation}
\begin{equation}
\begin{gathered}
W_{(3,3)}+3W_{(4,2)}+W_{(2,2,2)}-4W_{(3,2,1)}-3W_{(4,1,1)}+3W_{(2,2,1,1)}+7W_{(3,1,1,1)}-\\-10W_{(2,1,1,1,1)}+10W_{(1,1,1,1,1,1)}=0
\end{gathered}
\end{equation}
This identities correspond to $\hat{e}_1$ and $\hat{f}_1$ operators in (\ref{stabvir}).

\section{Discussion}

$\quad$We obtained the Virasoro constraints for two different cases. Resulting Virasoro generators are obviously connected. Such connection can be seen after applying them to the integral for torus knots with corresponding generating function(s) and putting all additional parameters equal to zero. Namely, we can make a following statement:
\begin{equation}
\left(\sum_{l=-\infty}^{\infty}(-1)^{l+H(l)}\left(\frac{\sqrt{q}}{1-q}\right)^{|l|}\frac{\partial}{\partial t_{nPQ+l}}-\frac{\partial}{\partial t_{nPQ}}\right)W\Bigg|_{t_k=0, \forall k}=-\frac{PQ}{\hat{g}_s}\frac{\partial}{\partial l_{nPQ}}\widetilde{W}\Bigg|_{t_k=l_k=0, \forall k}
\end{equation}
Perhaps it would look better if we rewrite this equality in terms of averages:
\begin{equation}
\Big\langle \sum_{l=-\infty}^{\infty}(-1)^{l+H(l)}\left(\frac{\sqrt{q}}{1-q}\right)^{|l|}x_i^{nPQ+l}-x_i^{nPQ}\Big\rangle=-\Big\langle\frac{PQ}{\hat{g}_s}\sum_{i=1}^Nx_i^{nPQ}\log(x_i) \Big\rangle
\end{equation}
Note that in the LHS we have replaced $z_i$ with $x_i$. This is a valid operation, as was discussed in section 3 (see (\ref{complex})). However, such relations are yet to be discussed. Direct application of Virasoro generators gives us identities between Wilson loops, or HOMFLY polynomials. These identities, contain either an undesirable term $\langle\sum x_i^{nPQ} \log(x_i) \rangle$ or infinite sum of Wilson loops in different representations that prevents us from getting closed recurrent relations. On the other hand, the logarithmic term can be directly computed, for example in $N=2$ case: 
\begin{equation}‹
\begin{split}
\langle\sum_i x_i^{nPQ} \log(x_i) \rangle = 8\hat{g}_s^2\pi e^{\frac{\hat{g}_s}{4}\left(\frac{(P-Q)^2}{PQ}+2n(PQ-P-Q)\right)}\Big(-e^{\hat{g}_s}(P+Q-2nPQ+1)+\\+e^{\hat{g}_s(1+n(P+Q))}(P+Q+2nPQ)+e^{\hat{g}_snQ}(P-Q-2nPQ)+e^{\hat{g}_snP}(Q-P-2nPQ)\Big)
\end{split}
\end{equation}
or it can be used to compare compositions of colored Wilson loops for two different knots, with equal products of $n$, $P$ and $Q$. The obtained Virasoro generators are also useful for studying these relations in the so-called stability limit, where they are significantly simplified.

\quad Though there is a working formula for HOMFLY polynomials or Wilson loops for all torus knots, the relations between these objects are not fully understood and this example of Virasoro algebra can give some idea of general structure of such algebras for all knots.

\section*{Aknowledgements}
\qquad Author is indebted to A. Gorsky for suggesting this problem and numerous fruitful discussions. I am grateful to A. Milekhin and Sh. Shakirov for useful comments and valuable insights. This work was supported in part by grant of Ministry of Education and Science of the Russian Federation under contract 8207, by grant RFBR-12-02-00284 and by grant NSh-3349.2012.2.

\end{document}